\documentclass{pasj01}
\draft 
\Received{$\langle$reception date$\rangle$}
\Accepted{$\langle$acception date$\rangle$}
\Published{$\langle$publication date$\rangle$}
\usepackage{xcolor}

\newcommand{\RCa}{black}
\newcommand{\RCb}{black}

\newcommand{\ADRS}{\textcolor{black}{ADRS} }
\newcommand{\ADRSns}{\textcolor{black}{ADRS}}

\begin{document}

\def\Herschel{{\it Herschel}}
\def\Spitzer{{\it Spitzer}}
\def\WISE{{\it WISE}}
\def\AKARI{{\it AKARI}}

\def\um{$\mu \rm{m}$}

\def\Xco{$X_{\rm CO}$}
\def\Xunit{$\rm{cm}^{-2}$ $\rm{K}^{-1}$ $\rm{km}^{-1}$ $\rm{s}$}
\def\cmcm{$\rm{cm}^{-2}$}
\def\cmcmcm{$\rm{cm}^{-3}$}
\def\kms{$\rm{km}$ $\rm{s}^{-1}$}
\def\Kkms{$\rm{K}$ $\rm{km}$ $\rm{s}^{-1}$}
\def\vlsr{$v_{\rm LSR}$}
\def\degree{$^{\circ}$}
\def\Lsun{$L_{\solar}$}
\def\Msun{$M_{\solar}$}
\def\Msunyr{$M_{\solar}$ $\rm{yr}^{-1}$ }

\def\NH{$N(\rm{H_{2}})$}

\def\HII{H \emissiontype{II}}
\def\OIII{O \emissiontype{III}}
\def\SII{S \emissiontype{II}}

\def\Ha{H$\alpha$}

\def\lb{($l$, $b$)}
\def\lbeq{($l$, $b$)$=$}
\def\lbsim{($l$, $b$)$\sim$}
\def\radec{($\alpha_{\rm J2000}$, $\delta_{\rm J2000}$)}
\def\radeceq{($\alpha_{\rm J2000}$, $\delta_{\rm J2000}$)$=$}
\def\radecsim{($\alpha_{\rm J2000}$, $\delta_{\rm J2000}$)$\sim$}

\def\COa{\atom{C}{}{12}\atom{O}{}{}}
\def\COb{\atom{C}{}{13}\atom{O}{}{}}
\def\COc{\atom{C}{}{}\atom{O}{}{18}}

\def\Jeq{{\it J}$=$}
\def\Ja{{\it J}$=1$--$0$}
\def\Jb{{\it J}$=2$--$1$}
\def\Jc{{\it J}$=3$--$2$}

\title{\textcolor{\RCa}{Observational demonstration of a} low-cost fast Fourier transform spectrometer with a delay-line-based ramp-compare ADC implemented on FPGA}

\author{Atsushi \textsc{nishimura}\altaffilmark{1*}}
\author{Takeru \textsc{matsumoto}\altaffilmark{1}}
\author{Teppei \textsc{yonetsu}\altaffilmark{1}}
\author{Yuka \textsc{nakao}\altaffilmark{1}}
\author{Shinji \textsc{fujita}\altaffilmark{1}}
\author{Hiroyuki \textsc{maezawa}\altaffilmark{1}}
\author{Toshikazu \textsc{onishi}\altaffilmark{1}}
\author{Hideo \textsc{ogawa}\altaffilmark{1}}

\altaffiltext{1}{Department of Physical Science, Graduate School of Science, Osaka Prefecture University, 1-1 Gakuen-cho, Naka-ku, Sakai, Osaka 599-8531, Japan}

\email{nishimura@p.s.osakafu-u.ac.jp}

%% `\KeyWords{}' always has to be placed before `\maketitle'.
\KeyWords{instrumentation: spectrographs} %Do NOT move this preamble from here!

\maketitle
\begin{abstract} % max 300 words.

In this study, a novel type of Fourier transform radio spectrometer (termed as
\textcolor{\RCa}{all-digital radio  spectrometer; ADRS}
) has been developed in which all functionalities comprising a radio spectrometer including a sampler and Fourier computing unit were implemented as a soft-core on a field-programmable gate array (FPGA).
A delay-line-based ramp-compare analog-to-digital converter (ADC), one of completely digital ADC, was used, and two primary elements of the ADC, an analog-to-time converter (ATC) and a time-to-digital converter (TDC), were implemented on the FPGA.
The sampling rate of the \ADRS $f$ and the quantization bit rate $n$ are limited by the relation, $\tau = \frac{1}{2^{n}f}$, where $\tau$ is the latency of the delay element of the delay-line.
Given that the typical latency of the delay element implemented on FPGAs is $\sim10$ ps, adoption of a low quantization bit rate, which satisfies the requirements for radio astronomy, facilitates the realization of a high sampling rate up to $\sim$100 GSa/s.
In addition, as the proposed \ADRS does not require a discrete ADC and can be implemented on mass-produced evaluation boards, its fabrication cost is much lower than that of conventional spectrometers.
The \ADRS prototype was fabricated with values of $f$ = 600 MSa/s and $n$ = 6.6 using a PYNQ-Z1 evaluation board, with a $\tau$ of 16.7 ps.
The performance of the prototype, including its linearity and stability, was measured, and a test observation was conducted using the Osaka Prefecture University 1.85-m mm-submm telescope; this confirmed the potential application of the prototype in authentic radio observations.
With 10 times better cost performance ($\sim$800 USD GHz$^{-1}$) than conventional radio spectrometers, the prototype facilitates cost-effective coverage of intermediate frequency (IF) bandwidths of $\sim100$ GHz in modern receiver systems.

\end{abstract}

\section{Introduction}

Heterodyne observation in the radio astronomy is a powerful technique for investigating interstellar medium (ISM) using frequency-related information detected from molecular emission lines.
Typically the technique provides a high spectral resolution of $F / \Delta F > 10^6$, where $F$ is a central frequency and $\Delta F$ is the frequency resolution of the system; therefore the technique is often used for detailed analysis of the dynamic states of ISM by resolving their Doppler velocity in $\sim 0.1$ \kms \ (e.g., \cite{2018PASJ...70S..42N, 2019PASJ..tmp...46F, 2020ApJ...896...36T}).
However, since the frequency coverage is typically limited to $BW / F \sim 0.05$, where $BW$ is the bandwidth observed by the system in parallel, the technique is inefficient for wider bandwidth observations, including multi-line observations and unbiased surveys for detection of red-shift.
Recently, wide-bandwidth heterodyne receivers for detection of radio frequency (RF) and intermediate frequency (IF) have been developed with an output bandwidth of $\sim20$ GHz \citep{2018.ieee.kojima, 2020A&A...640L...9K, 2020SPIE.masui, 2020SPIE.yamasaki}.
Moreover, to extend the survey coverage of molecular line mappings toward expansively distributed giant molecular clouds, multibeam receivers have been created using wide-bandwidth technologies \citep{2016SPIE.9914E..1ZM, 2020SPIE.nishimura.nasco}. 
These multibeam receivers can now attain total IF output bandwidths of $\sim100$ GHz; however, this expansion requires development of associated back-end systems with wider bandwidths and lower costs.

% 分光計の話
Improvements in the bandwidths of digital radio spectrometers have been possible because of the development of enhanced computing techniques. 
Typically, radio spectrometers comprise two parts: (i) a sampler that digitizes the voltage of the input radio signal and (ii) a calculator that computes the power spectrum from the sampled data.
Consequently, achieving a wider spectrometer bandwidth requires faster samplers and more powerful calculators.
The AC240 fabricated by Acquiris was the first digital spectrometer with a bandwidth of 1 GHz to be commercially developed for applications in radio astronomy \citep{2005A&A...442..767B}.
During the following decade, several spectrometers with bandwidths of several GHz are developed \citep{2006A&A...454L..29K, 2012A&A...542L...3K, 2012PASJ...64...29K, 2013.balloon.stachnik, 2014PASP..126..761J, 2016PASP..128k5002J, 2016JAI.....541001H, 2017EP&S...69...95I, 2020PASP..132h5001J}, all of which possessed a discrete analog-to-digital converter (ADC) for the sampler and a field programmable gate array (FPGA) for the calculator.
Even with a narrow back-end bandwidth, the total bandwidth of the back-end system can be increased using multiple back-ends connected by an adequate IF circuit.
Thus, the cost of the spectrometer is an essential consideration in realizing a wideband back-end system.
To reduce the cost of spectrometers, graphics processing unit (GPU) spectrometers that use GPUs instead of FPGAs for the calculation of fast Fourier transforms (FFTs) are proposed (e.g., \cite{2014JAI.....350010M, 2014PASA...31...48C, 2019adw..confE..35M}).

\begin{figure}[t]
 \begin{center}
  \includegraphics[width=\textwidth]{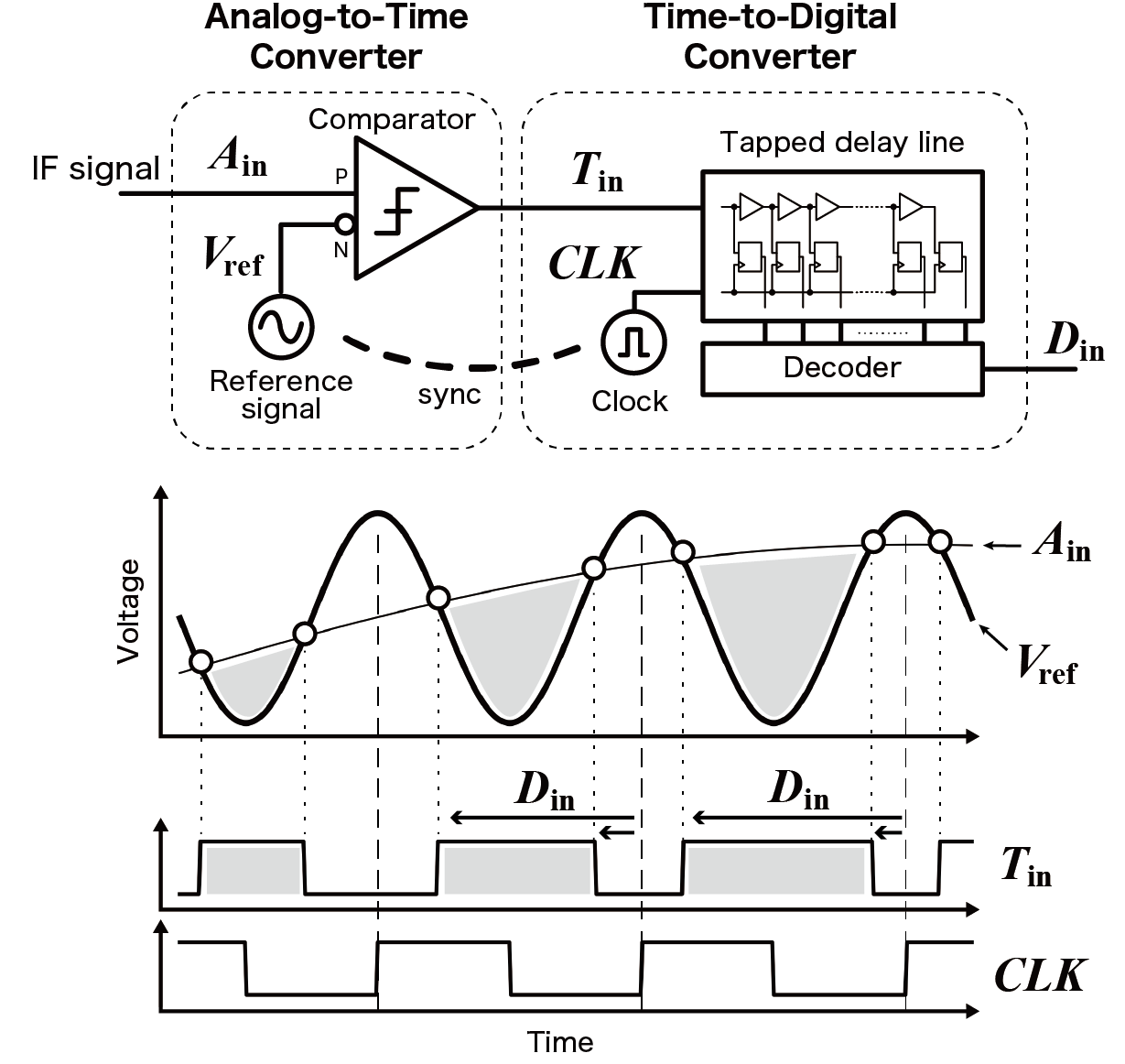}
 \end{center}
\caption{
Upper: Architecture of the delay-line-based ramp-compare type analog-to-digital converter (ADC) comprising an analog-to-time converter (ATC) and a time-to-digital converter (TDC).
Lower: Timing diagram of the ADC. 
$A_{\rm in}$ indicates the analog voltage input to the ATC, and $V_{\rm ref}$ indicates the reference signal used for the voltage comparison in the ATC.
$T_{\rm in}$ shows the output signal of the ATC that is input to the TDC, and $CLK$ is the clock signal used in the TDC.
$D_{\rm in}$ is calculated by detecting the positive and negative edges of the $T_{\rm in}$ signal at the decoder.
The phases of $V_{\rm ref}$ and $CLK$ are synchronized during the operation.
}
\label{fig:adc}
\end{figure}

% ramp-compare ADC の話
This paper addresses this need for cost reduction by implementing all the functionalities of radio spectrometers within a single FPGA chip (hereafter, \textcolor{\RCa}{all-digital radio spectrometer; ADRS}).
To implement an ADC function on an FPGA as a soft-core, this study adopts a delay-line-based ramp-compare fully digital ADC \citep{2004.tdcadc.kamas, 2009.ieee.li}.
The ADC mainly comprises two parts, an analog-to-time converter (ATC) and a time-to-digital converter (TDC), and all of the components constituting the ATC and TDC can be implemented on FPGAs \citep{2007.IEEE.wu, 2013EPJA...49...25E, 2013arXiv1311.6127P, 2015.homulle, 2016.ieee.homulle, 2018.ieee.xiang}.
Figure \ref{fig:adc} presents the principle of operation of the ramp-compare ADC with a delay-line.
The ATC is implemented using the continually repeating reference signal (e.g., sine wave) and the comparator, which compares the voltages of the reference and the input IF signals.
The delay-line type TDC \citep{1984.tdc.hoppe, 2019.ieee.tanncock} is used for the ADC.
% Section 2.1 provides the details of the principle of operation.
Although ADCs with a sampling rate of over 1 GHz \citep{2016.ieee.homulle}, multiple input channels \citep{2018.ieee.xiang}, and working under the ultralow temperature \citep{2016.ieee.homulle, 2018.ieee.xiang} have been realized on FPGAs till date, the implementation of radio spectrometers using the ADC has not been reported.
In this paper, a radio spectrometer equipped with a delay-line-based ramp-compare type ADC has been developed to examine the feasibility for authentic radio observations.
The sampling rate was 600 MSa/s with a quantization bit rate of 6.6.
The ADC and an FFT core were implemented on an FPGA (XC7Z020-1CLG400C, Xilinx) mounted on an evaluation board (PYNQ-Z1, Digilent).
Section 2 presents the architecture of the developed spectrometer including the operating principle of the ramp-compare ADC, its implementation within the FPGA, and the design of the FFT calculation.
\textcolor{\RCb}{
Section 2.1 provides the details of the principle of operation.
}
Section 3 provides descriptions of the calibration of the spectrometer, and Section 4 presents the results of the performance evaluation and the test observations.
Section 5 contains the summary of the paper.

\section{System architectures}

\begin{figure}[t]
 \begin{center}
  \includegraphics[width=\textwidth]{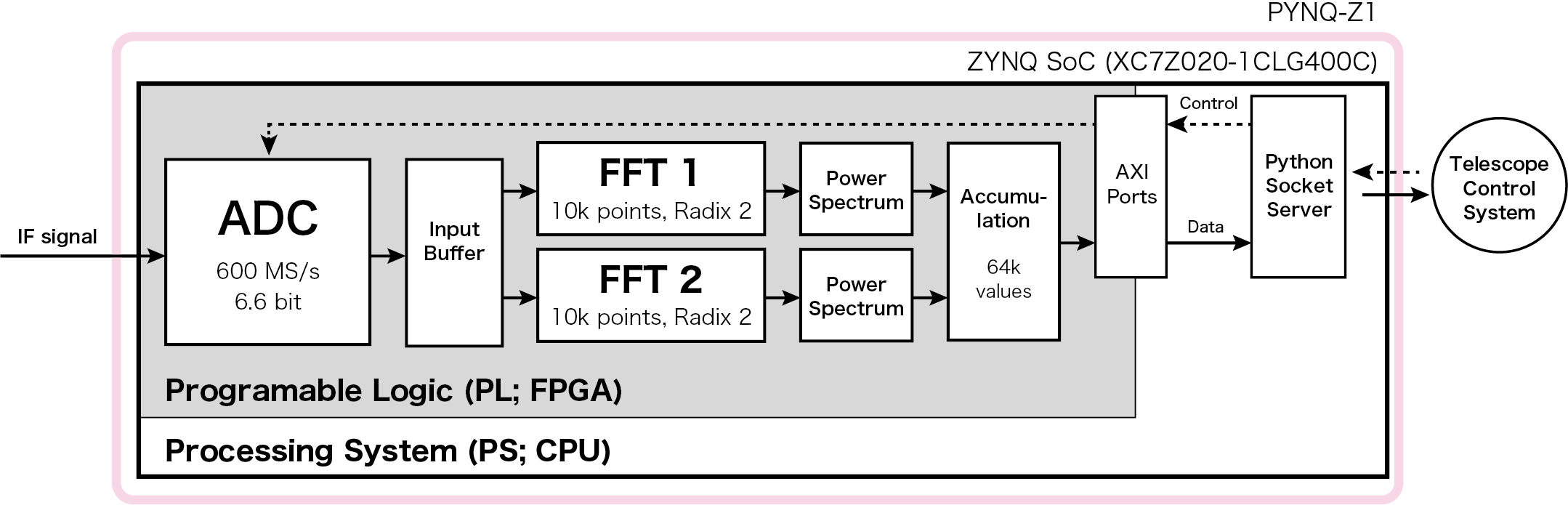}
 \end{center}
\caption{
Architecture of the prototype of the \textcolor{\RCa}{all-digital radio} spectrometer (\ADRSns) with a sampling frequency of 600 MSa/s.
The solid arrows indicate the flow of data for the processed intermediate frequency (IF) signal, and the dashed arrows indicate the flow of the analog-to-digital converter (ADC) control.
}
\label{fig:overall}
\end{figure}

\begin{figure}[t]
 \begin{center}
  \includegraphics[width=\textwidth]{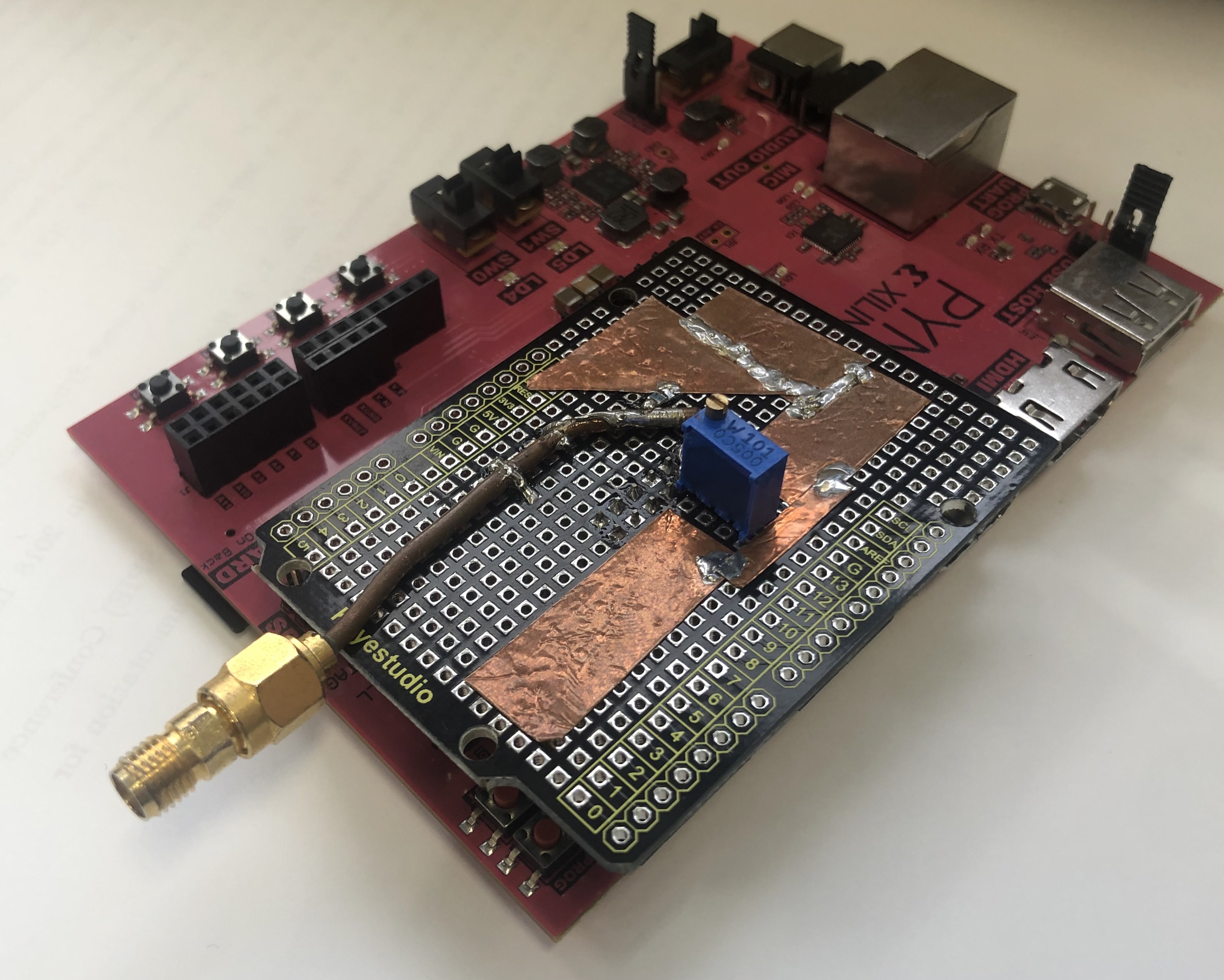}
 \end{center}
\caption{
Photograph of the prototype of the \textcolor{\RCa}{all-digital radio} spectrometer.
}
\label{fig:photo}
\end{figure}

A Digilent PYNQ-Z1 board was used for the \ADRS demonstrations.
This board was selected for the prototype because of its low cost, easy handling by the Linux OS, and its provision of a well-prepared PYNQ software framework.
The PYNQ-Z1 board contains an FPGA chip (ZYNQ XC7Z020-1CLG400C, Xilinx), which is a programmable system-on-a-chip (SoC) containing both an FPGA and a central processing unit (CPU) that can run the Linux OS.
PYNQ is a Python-based open-source framework that provides accesses to the FPGA controlling and communicating with the logics implemented on the FPGA.
Figure \ref{fig:overall} illustrates the overall architectures of the 600 MSa/s \ADRS implemented on the PYNQ-Z1 board.
In the \ADRSns, both ADC and FFT are implemented within the FPGA.
The accumulated power spectra are downloaded to the Python server via an advanced extensible interface (AXI) port connection, and the network delivers the data to the telescope control system.
Figure \ref{fig:photo} presents a photograph of the \ADRS prototype.
The dimensions of the prototype are 88 mm $\times$ 124 mm, and its typical power consumption under ADC and FFT operation is $\sim4$ W.
Its specific architectures are described in the following subsections.

\subsection{Delay-line-based ramp-compare analog-to-digital converter core}

Figure \ref{fig:adc} presents the operating principle of the delay-line-based ramp-compare type ADC used in this study.
The basic concept of the converter is the same as that used by \citet{2015.homulle}.
There are two main components in the ADC: the ATC and the TDC.
The ATC is contains a comparator and a reference signal generator.
The low voltage differential signaling (LVDS) element is used for the comparator by disabling its termination resistor.
The analog input signal, $A_{\rm in}$, is connected to the positive port of the LVDS.
The output of the mixed-mode clock manager (MMCM) primitive is used for the reference signal, $V_{\rm ref}$.
Because the MMCM output is a square wave, it is first output via general-purpose input-output (GPIO) connector of the PYNQ-Z1 board before passing it through an RC filter implemented on the interface board to generate the triangular wave; thereafter, the signal is directed to the negative port of the LVDS element.
The comparator outputs the signal, $T_{\rm in}$, of the high-level (or low-level) voltage when the voltage of the positive port is higher (or lower) than that of the negative port.
The TDC comprises a delay-line and a decoder.
The delay-line monitors two pulses, $T_{\rm in}$ and a clock signal $CLK$, and detects the time offset between the negative and positive edges of the $T_{\rm in}$ pulse and the positive edge of the $CLK$ pulse.
The phase calibration \textcolor{\RCb}{described} in Section \ref{sec:calib-phase} synchronizes the phases of the $V_{\rm ref}$ and $CLK$ during operation.
Assuming that the time resolution of the TDC is $\tau$, the sampling frequency, $f$, and the quantization bit rate, $n$, are restricted by the following relation (see also Eq. 1 in \cite{2009.ieee.li}):
\begin{equation}
\tau = \frac{1}{2^nf}.
\end{equation}
The delay-line is implemented using the carry connections as delay elements and flip-flops (FFs) to maintain the condition of the $T_{\rm in}$ signal.
The delay time, $\tau$, of each carry element for the XC7Z020-1CLG400C is 16.67 ps, and a maximum of 200 chains can be implemented within the same clock domain in the FPGA.
Thus, in theory, an ADC with ($f$, $n$)=(600 MSa/s, 6.6 bit) and (60 GSa/s, 1 bit) can be realized on the FPGA chip.
This paper reports a prototype with a sampling rate of $f=600$ MSa/s, whereas another prototype with $f=60$ GSa/s will be reported in a separate paper (Matsumoto et al. inprep.).
A delay-line with 200 elements was implemented.
In the decoder, the output signal from the delay-line (0--199 bits) is first separated to the anterior (0--99 bits) and posterior (100--199 bits) halves containing the positive and negative edges of the $T_{\rm in}$, respectively, and then, the numbers of high-level bits in each of 100 bits signals of the TDC code are counted using look-up tables (LUTs).
The decoder converts the detected TDC code into a digitized input voltage, $D_{\rm in}$, according to the calibration table, which is measured before operation (see Section \ref{sec:calib-amp}).
A 300 MHz clock drives both the ATC and the TDC.
The TDC outputs the positive and negative edges of the $T_{\rm in}$ signal as two sampling values, which provides an ADC sampling rate of 600 MSa/s.

\subsection{FFT core}

This study used Radix-2 algorithm with the decimation-in-time (DIT) method to program and implement a real-time 1,024 point FFT core.
Two FFT cores operating with a 300 MHz clock were used to calculate the ADC code of 600 MSa/s.
Because the DSP48E1 digital signal processing slice can calculate the multiplication of the integers of 25 and 18 bits within one clock, 25 bits were assigned for the input and temporary values during the FFT calculations with 18 bits assigned for the twiddle factor.
The values were processed as fixed-point numbers, and the twiddle factors for each FFT calculation were stored in the block random access memories (BRAMs).
A series of FFT calculations provided a power spectrum of 512 frequency points.
Thereafter, the power spectra were accumulated 65,536 times corresponding to an integration time of 111.8 ms (see also Figure \ref{fig:overall}).
Finally, the integrated power spectrum was stored in the BRAM, which was accessible from the CPU via the AXI connections.
% Figure \ref{fig:fpga} shows the implementation of the ADC and FFT cores within the FPGA.
Since the prototype demonstrates a utilization rate of FPGA elements that is below 36\%, the PYNQ-Z1 boards can be used to implement a spectrometer with spectral channels exceeding 1,024 points.

\subsection{Data acquisition server}

The Python-based socket server program has been developed to provide the power spectra to the telescope control system.
The integrated power spectrum stored in the BRAM was downloaded to the CPU via AXI ports, using the Python library {\tt pynq}\footnote{https://pynq.readthedocs.io/}.
The ADC calibrations and data download from the \ADRS were controlled via the server program from socket clients such as telescope control systems.

\section{System calibrations}

The ADC requires two calibrations, corresponding to time and voltage domains, before its operation.
\citet{2016.ieee.homulle} summarized the basic concept of the calibration procedures.
The following subsections describe the calibration procedures and the results for the \ADRS prototype.

\subsection{Phase calibration}
\label{sec:calib-phase}

\begin{figure}[t]
 \begin{center}
  \includegraphics[width=\textwidth]{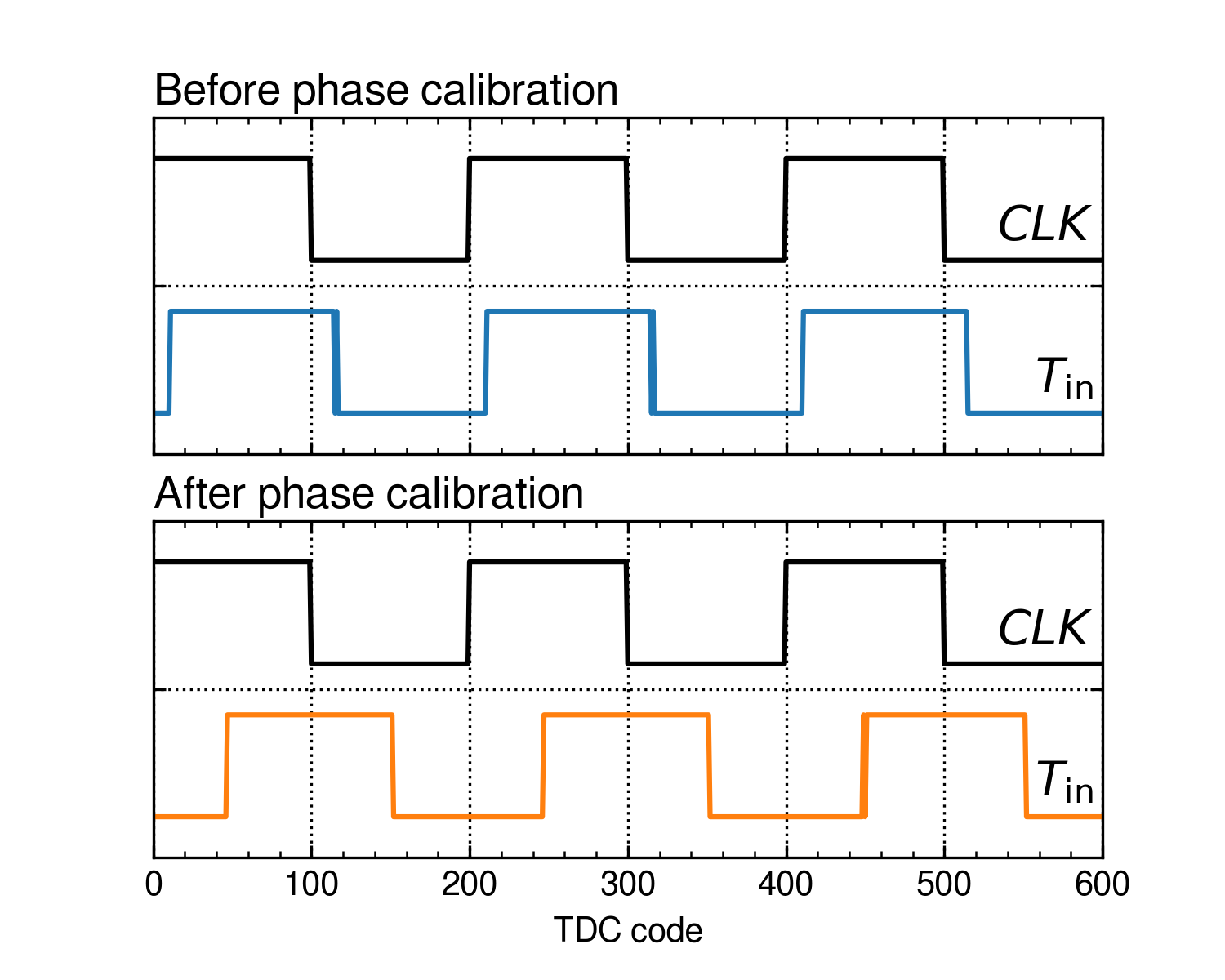}
 \end{center}
\caption{
Timing chart of the $T_{\rm in}$ signal output from the time-to-digital converter (TDC).
The upper and lower figures show the examples before and after the calibration.
}
\label{fig:calib-phase}
\end{figure}

The ramp-compare ADC requires the phases of the reference signal in the ATC and the clock in the TDC to be \textcolor{\RCb}{synchronized}.
The ADC implemented in this study used the reference signal, $V_{\rm ref}$, which is generated by the MMCM in the FPGA and is shaped as a triangular wave through the RC filter implemented in the external circuit, before being returned to the LVDS input of the FPGA. 
On the other hand, the clock signal, $CLK$, is generated by the same MMCM and internally connected to the TDC.
Hence, significant differences exist between the paths of the $V_{\rm ref}$ and $CLK$, and the phase of the $V_{\rm ref}$ must be aligned via the calibration process.
This study implemented the programmable delay element, IDELAYE2, within the wire of the $T_{\rm in}$ signal between the ATC and TDC.
The signal input to the IDELAYE2 primitive is output by a delay  of 0--1,600 ps, and during operation, this delay can be programmed in 32 levels.
Figure \ref{fig:calib-phase} contains the timing chart of the $T_{\rm in}$ signal before and after calibration.
Here, the input bias, $A_{\rm in}$, of 0 V is used.
The $T_{\rm in}$ signal is aligned after calibration, and the positive and negative edges are seen to be located at the center of the anterior and posterior halves corresponding to TDC counts of 50 and 150, respectively.

\subsection{Amplitude calibration}
\label{sec:calib-amp}

\begin{figure}[t]
 \begin{center}
  \includegraphics[width=\textwidth]{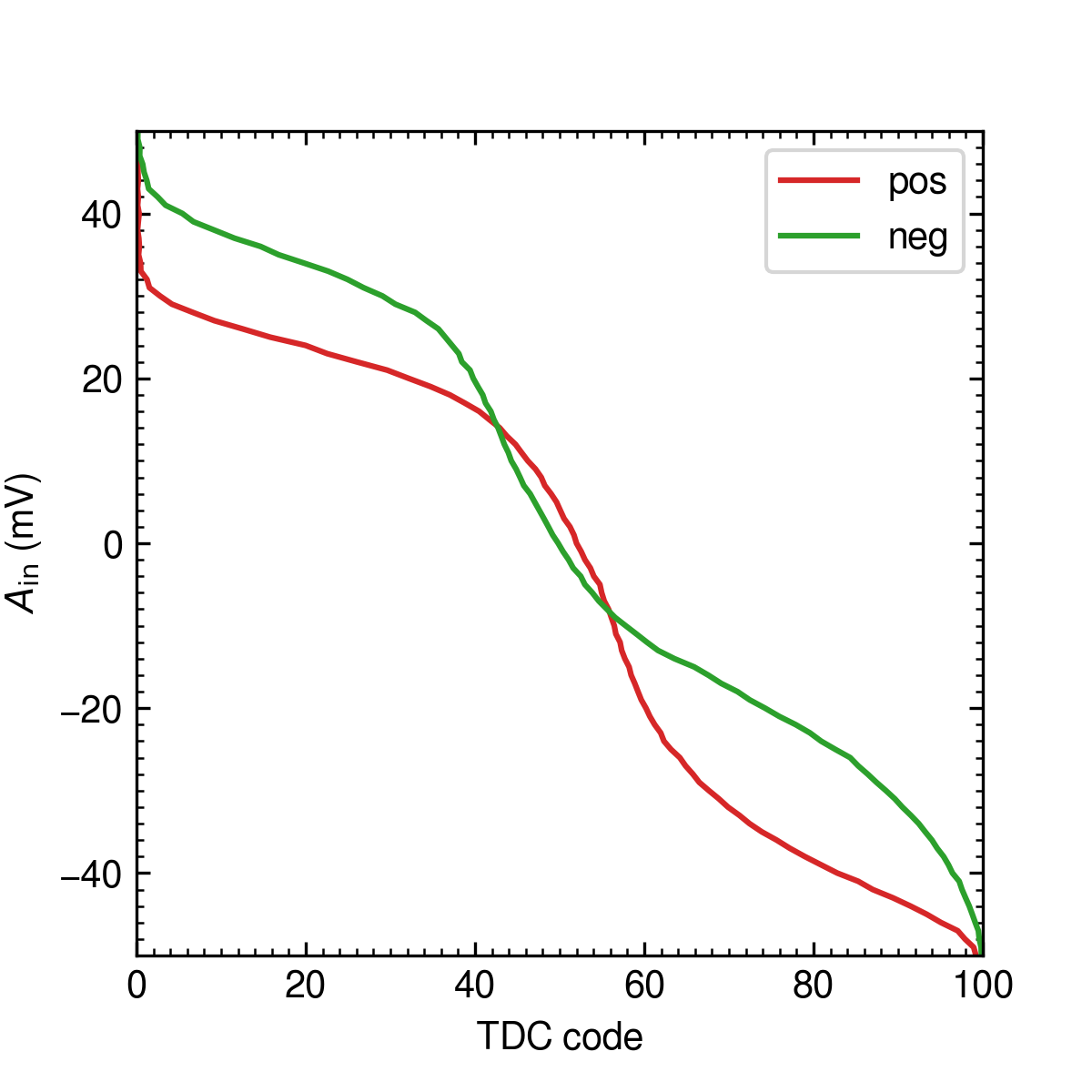}
 \end{center}
\caption{
Graph of the conversion relation from the time-to-digital converter (TDC) code, $T_{\rm in}$, to the input bias voltage, $A_{\rm in}$.
The calibration curves for the positive and negative edges of the $T_{\rm in}$ signals are indicated in red and green colors respectively.
}
\label{fig:calib-amp}
\end{figure}

\begin{figure}[t]
 \begin{center}
  \includegraphics[width=\textwidth]{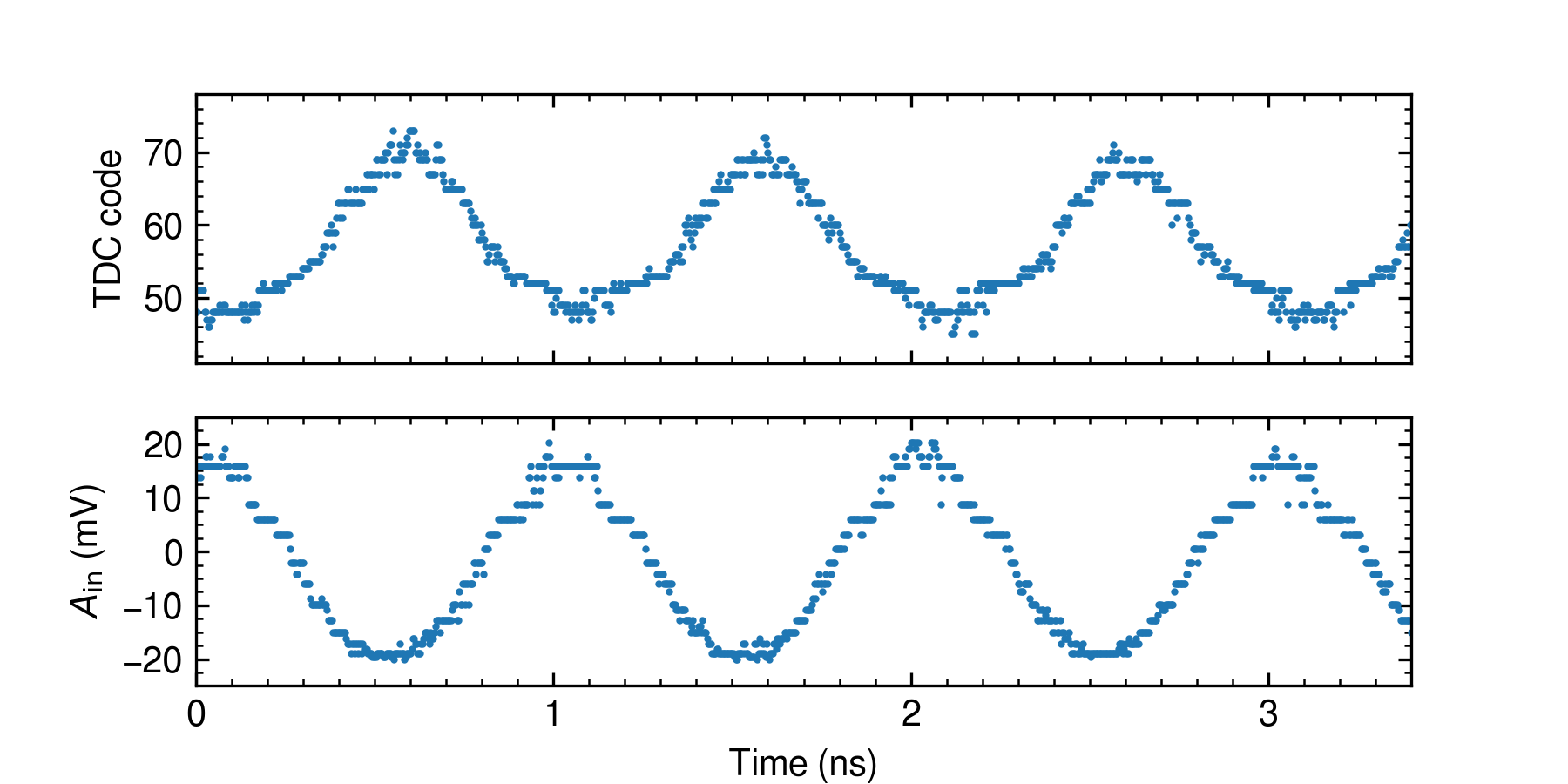}
 \end{center}
\caption{
Time-to-digital (TDC) code and input voltage ($A_{\rm in}$) digitized by the \textcolor{\RCa}{all-digital radio} spectrometer (\ADRSns) prototype for a sinusoidal period of 1 MHz.
}
\label{fig:data-adc}
\end{figure}

The reference signal, $V_{\rm ref}$, is not an ideal triangular wave; thus, its actual shape must be measured and corrected appropriately.
The shape of the $V_{\rm ref}$ is measured by applying a constant voltage to $A_{\rm in}$ and measuring the TDC code, $T_{\rm in}$.
The results of the calibration measurements are provided in Figure \ref{fig:calib-amp} and indicate the conversion relation from the TDC code, $T_{\rm in}$, to the input bias, $A_{\rm in}$.
Expectedly, the conversion curve of the ideal and most efficient triangular wave will be a straight line; however, the line produced by the \ADRS prototype is wavy.
The error from the ideal straight line is greater in the \ADRS prototype proposed in this paper than in that reported by \citet{2016.ieee.homulle}, although both architectures are almost identical.
This is regarded to be due to that originally, the transmission lines used for the reference signal of the \ADRS prototype were designed for digital communications and were not intended for high-frequency analog signal handling purposes.
This could be enhanced by re-designing outer circuit considering the actual GPIO line parameters, such as parasitic resistance and parasitic conductance.
Broadly, the gap from the straight line causes the nonlinearity of the ADC, as normal ADCs assume an output ADC code of equally spaced interval values.
To avoid this nonlinearity, this paper adopted fixed-point numbers for the ADC output code, $D_{\rm in}$.
The measured $T_{\rm in}$ value was decoded to the $D_{\rm in}$ of the signed fixed-point value with a bit number of 25; this length is identical to that used in the FFT calculations.
The measured conversion table was stored in the BRAM, and the decoding was operated in one clock.
Although the output of the ADC code is not an equally spaced interval value when using this configuration; in principle, it produces an ADC code that is free of systematic errors, including gain error, offset error, and integral nonlinearity (INL).
Figure \ref{fig:data-adc} shows the measured TDC code, $T_{\rm in}$, and the ADC code, $D_{\rm in}$.

Fluctuations in the operating temperature of the FPGA are considered to cause changes in the calibration parameters described above.
The resulting effects on the accuracy of the power and frequency of the obtained spectrum necessitate further investigation.

\section{Evaluation of performance}

\subsection{General performance}

\citet{2015.homulle} and \citet{2016.ieee.homulle} investigated the performance of a delay-line-based ramp-compare ADC implemented on an FPGA.
This subsection describes the results of the performance evaluations of the \ADRS prototype particularly in the context of its application in radio astronomy observations.

\begin{figure}[t]
 \begin{center}
  \includegraphics[width=\textwidth]{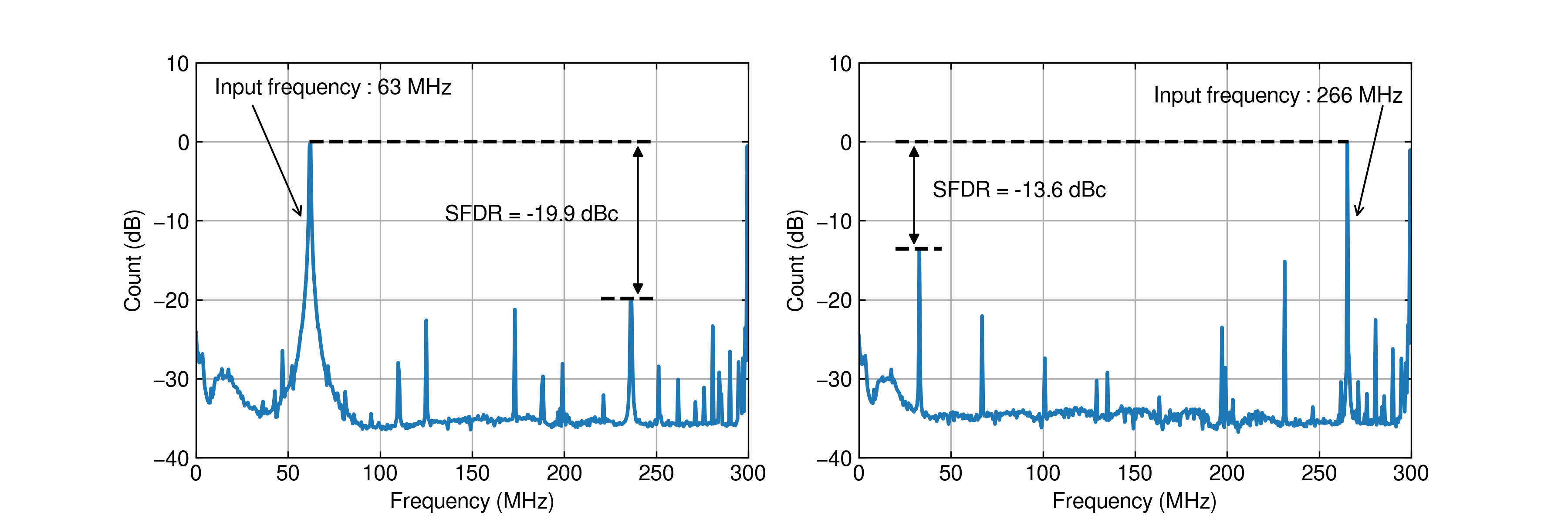}
 \end{center}
\caption{
Measured spurious-free dynamic range (SFDR) of the \textcolor{\RCa}{all-digital radio} spectrometer (\ADRSns) prototype for sinusoids of 63 MHz (left) and 266 MHz (right).
}
\label{fig:adc-snr}
\end{figure}

% SFDR, SNDR, ENOB
Figure \ref{fig:adc-snr} illustrates the plots of the power spectra measured by the \ADRS prototype.
The signals of continuous sine waves generated by the signal generator at frequencies of 63 and 266 MHz were input to the \ADRSns.
Spurious-free dynamic ranges (SFDRs) of $-19.9$ and $-13.6$ dBc were measured for input frequencies of 63 and 266 MHz, respectively, while the same frequencies generated signal-to-noise-and-distortion ratios (SNDRs) of $-10.0$ and $-5.4$ dB respectively.
The effective number of bits (ENOB) was calculated to be 1.4 and 0.6 for input frequencies of 66 and 266 MHz, respectively, using the following equation:
\begin{equation}
ENOB = \frac{SNDR - 1.76}{6.02}.
\end{equation}
% 結果の解釈を入れる
\textcolor{\RCa}{
In general, the ENOB of the delay-line-based ramp-compare ADC is low especially for the higher frequency side. 
\citet{2016.ieee.homulle} reported that the typical ENOB of their 1.2 GSa/s 6.6 bit ADC at an input frequency of 100 MHz is $\sim$3.
An ENOB is lower if the noise floor level is higher or the level of spurious signal is higher.
For the \ADRS prototype, the input analog signal is sampled with an unequally spaced time and is quantized with an unequally spaced voltage, and these two factors cause the increase of the qantization error in the output ADC code.
The level of spurious signal is increased if the quantization error is correlated to the time, while in other case the noise floor level is increased.
}
In this paper, the level of the noise floor and the spurious level of the \ADRS prototype are both higher than in the model used by \citet{2016.ieee.homulle}; consequently, the ENOB of the \ADRS prototype is lower than that reported by \citet{2016.ieee.homulle}.
\textcolor{\RCa}{
This is mainly due to the non-ideality of the reference signal (see Figure \ref{fig:calib-amp}).
However, the low ENOB of the \ADRS seems not to affect the linearity dynamic range of input power critically (see below).
Furthermore, the system noise temperature and the antenna temperature of the celestial object are correctly measured by the \ADRS prototype (see Section \ref{sec:test-obs}), suggesting that the low ENOB does not affect the accuracy of the radio observations.
}
% The low ENOB may affect the observational efficiency, and this topic should be addressed in future research.

\begin{figure}[t]
 \begin{center}
  \includegraphics[width=\textwidth]{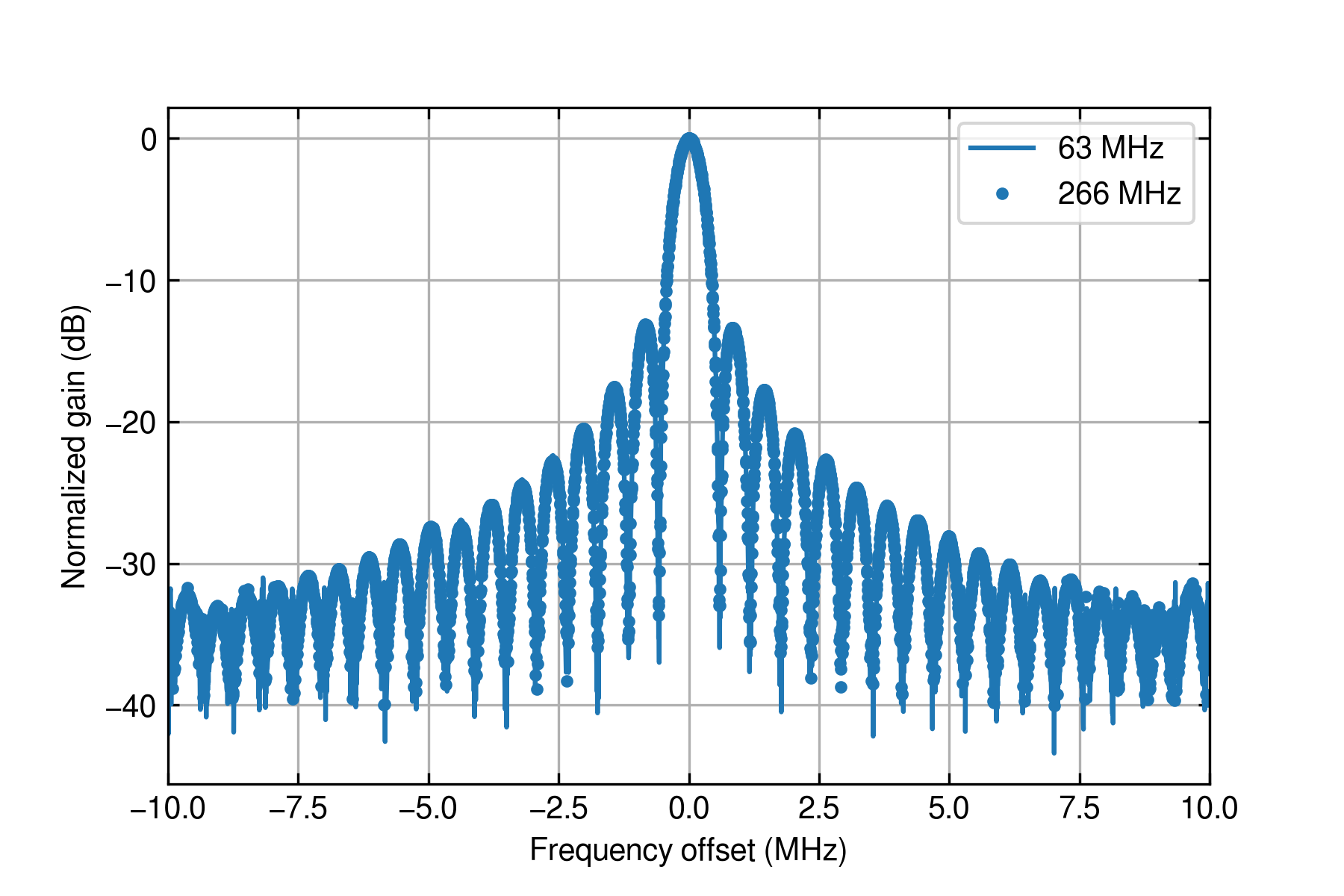}
 \end{center}
\caption{
Frequency response of the \textcolor{\RCa}{all-digital radio} spectrometer (\ADRSns) prototype for frequencies of 63 and 266 MHz.
The rectangular window function is used.
}
\label{fig:adc-freqres}
\end{figure}

% 周波数応答
Figure \ref{fig:adc-freqres} shows the measured frequency response of the \ADRS prototype.
The full width at half maximum (FWHM) of the main lobe was measured to be 515.8 kHz, which corresponds to a bin width of 0.88, and the level of the side lobe was measured to be $-13.1$ dB.
No window functions were implemented for the \ADRS prototype. 
These performances are consistent with the frequency response for the rectangular window, whereas the implementation of window functions or digital pre-filters such as polyphase filter bank is an area for future study.
The response curves were almost identical over the entire 300 MHz bandwidth of the \ADRS prototype, and it was confirmed that the the frequency response of the spectrometer demonstrated no negative effects when the delay-line-based ramp-compare ADC was adopted.

\begin{figure}[t]
 \begin{center}
  \includegraphics[width=\textwidth]{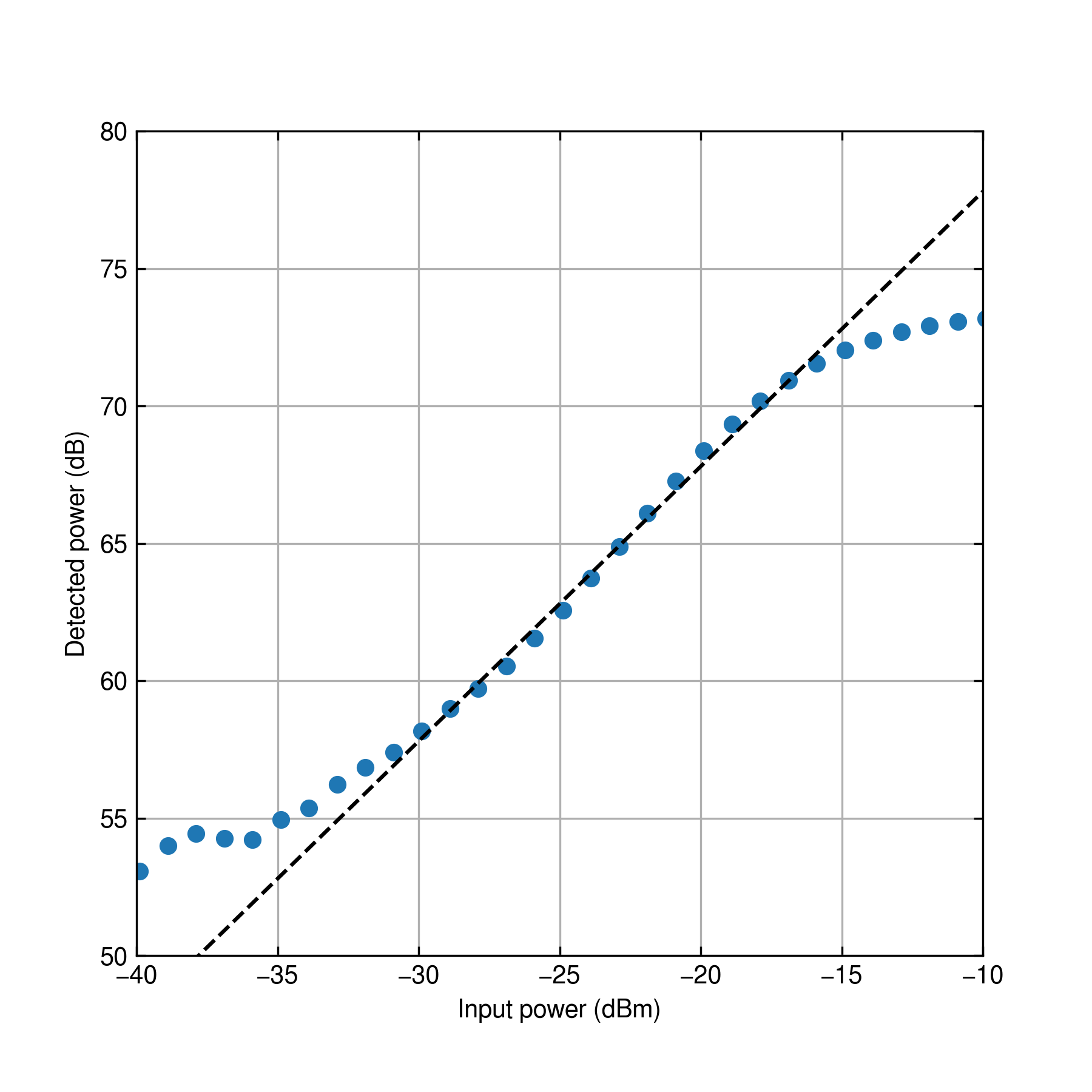}
 \end{center}
\caption{
Relationship of the input power and the detected power levels of the \textcolor{\RCa}{all-digital radio} spectrometer (\ADRSns) prototype.
The dashed line indicates the ideal linear response for reference.
}
\label{fig:adc-lin}
\end{figure}

% 線形性
% 意義
Figure \ref{fig:adc-lin} presents the results of linearity evaluation for the detected power of the \ADRSns, which were obtained by measuring the spectra of a laboratory noise source under various power inputs.
The detected power demonstrated a linear response for an input power range of $-30$--$-17$ dBm with an error below 10\%.
This dynamic range of 13 dB is wide enough for heterodyne observations that require a dynamic range of typically $\sim5$ dB, according to the power ratio of a source and calibrator operating at room temperature.

\begin{figure}[t]
 \begin{center}
  \includegraphics[width=\textwidth]{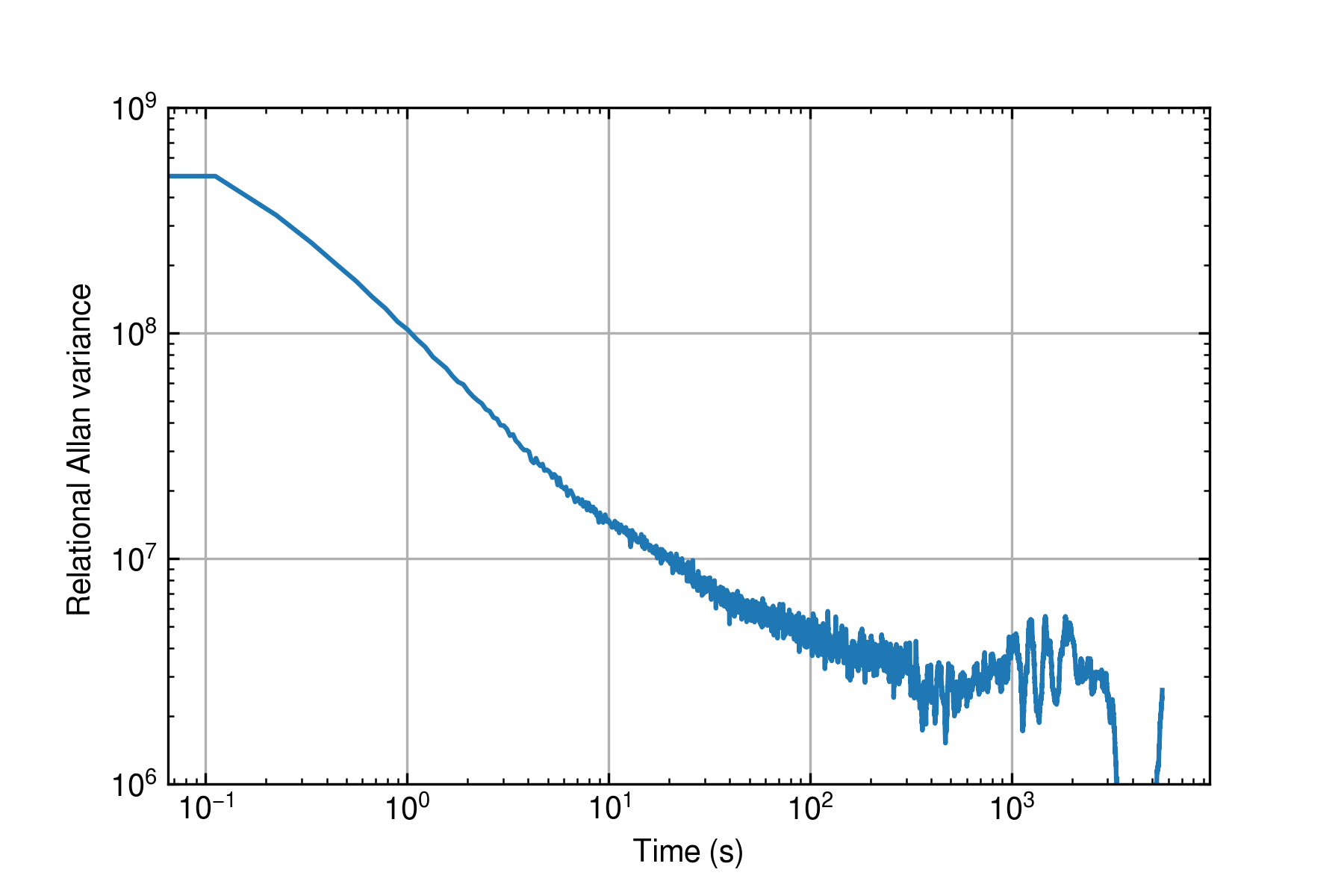}
 \end{center}
\caption{
Relational Allan variance plot of the \textcolor{\RCa}{all-digital radio} spectrometer (\ADRSns) prototype.
The data with a frequency separation of the 15 MHz was used.
}
\label{fig:adc-allan}
\end{figure}

% アラン分散
% 意義
The spectroscopic Allan variance was evaluated by measuring and processing the spectrum of the noise source, resulting in an Allan time of $>100$ s (Figure \ref{fig:adc-allan}).
A frequency separation of 15 MHz was used for the analysis.
During the measurements, the room temperature fluctuated by $\sim 1$ K, which is similar to the observation condition at the location of the 1.85-m radio telescope.

\subsection{Test observations}
\label{sec:test-obs}
% rms の軽減率を入れる
% スペクトルの解釈を追加

\begin{figure}[t]
 \begin{center}
  \includegraphics[width=\textwidth]{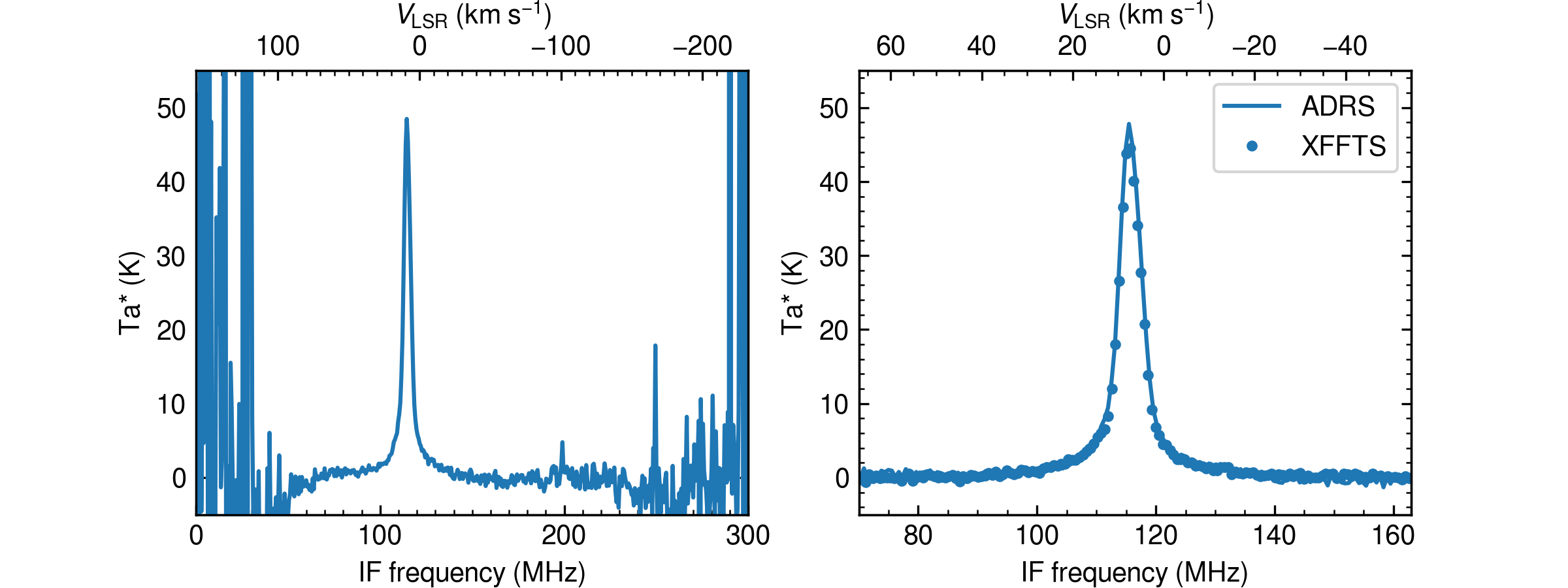}
 \end{center}
\caption{
Observed spectrum of $^{12}$C$^{16}$O $J=2-1$ toward Ori-KL.
Left: the entire the spectrum obtained by the \textcolor{\RCa}{all-digital radio} spectrometer (\ADRSns) prototype.
Right: zoom-up view of the $^{12}$C$^{16}$O $J=2-1$ line. The line and dots indicate the spectrum obtained by the \ADRS and XFFTS, respectively.
}
\label{fig:obs}
\end{figure}

First-light observations were conducted in December 2020 using the Osaka Prefecture University 1.85-m mm-submm telescope \citep{2013PASJ...65...78O, 2020SPIE.nishimura.1p85} with a 230 GHz band receiver \citep{2020SPIE.masui, 2020SPIE.yamasaki} installed at Nobeyama Radio Observatory.
The receiver had an output bandwidth of 2 GHz, and a low-pass-filter (LPF), Mini-Circuits VLF-180 with a bandwidth of 180 MHz, was inserted prior to the \ADRS prototype.
The frequency interval of the spectrum obtained by the \ADRS prototype was 585.9 kHz corresponding to the velocity resolution of 0.76 km s$^{-1}$ for the 230 GHz band.
The IF output from the receiver was simultaneously recorded by RPG XFFTS \citep{2012A&A...542L...3K}, and the observation results were compared.
The system noise temperature including atmosphere, $T_{\rm sys}$, was measured  by the \ADRS prototype to be $\sim600$ K, which represents a difference of $<10\%$ to the XFFTS value.
The weather conditions during the investigation were cloudy.
Figure \ref{fig:obs} shows the spectrum of the $^{12}$C$^{16}$O $J=2-1$ line toward Ori-KL obtained by position switching observations with an integration time of 180 s.
After subtracting the baseline, the peak temperature of the spectrum was the same as the XFFTS result within an error of 10\%, and the shape of the spectrum is consistent with the XFFTS result.
However, the observation efficiency, $\varepsilon$, of the \ADRS prototype is 20\% lower than that of the XFFTS, where $\varepsilon$ is defined by the following equation:
\begin{equation}
\varepsilon = \frac{T_{\rm sys}}{T_{\rm rms} \sqrt{dt \cdot df}},
\end{equation}
where, $T_{\rm rms}$ is the rms of the spectrum, $dt$ is the integration time, and $df$ is the frequency resolution.
The lower efficiency of the \ADRS prototype is considered to be mainly attributed to environmental noise emanating from the telescope instruments.
Because of fabrication deficiencies in the shield of the prototype \ADRSns, its ground-level position was affected by environmental noise from the servomotors and cryocooler of the receiver. 
From this observation, the basic concept of the \ADRS was confirmed to be feasible for application in the field of radio astronomy; however, improvements in noise immunity are necessary to enable efficient observations comparable with those obtained from commercial spectrometers.

\section{Summary}

This study developed an \ADRSns, which is a novel radio spectrometer based on a single FPGA chip that implements delay-line-based ramp-compare ADC and FFT as soft-core.
The operating principle of the ADC and the implementations of the ADC and FFT soft-core on the PYNQ-Z1 evaluation board were presented in detail.
The first \ADRS prototype-based radio spectrometer was developed with a sampling rate of 600 MHz and a quantization bit rate of 6.6.
Following investigations into its performance, the stability of the \ADRS for application in astronomical observations has been confirmed.
Experimental validation was performed by observing the $^{12}$C$^{16}$O $J=2-1$ line toward a star forming molecular cloud using the 1.85-m radio telescope with a 230 GHz band receiver.
The obtained spectrum produced a shape that was consistent with that obtained by the XFFTS simultaneously within an error of $<10$\%.
Because of the low cost of the PYNQ-Z1, the cost performance of the prototype is around $800$ USD GHz$^{-1}$, which is more than 10 times lower than that of other radio spectrometers.
Thus, the economic and performance feasibility of implementing the \ADRS to cover all IF output from modern receiver systems such as FOREST, which is achieving a bandwidth of $\sim100$ GHz, has been confirmed.

In principle, increasing the sampling rate, $f$, of the delay-line-based ramp-compare ADC by reducing the quantization bit rate, $n$, allows the extreme solution of ($f$, $n$)$=$(60 GSa/s, 1 bit) to be implemented using a PYNQ-Z1 board. 
Furthermore, using shorter delay elements to implement the TDC could result in an \ADRS with an even higher sampling rated.
The XC7Z020-1CLG400C FPGA used in this paper was created in a 28 nm process, whereas some of the more modern FPGAs involve 16 nm processes.
It is considered that the delay time of the carry chain element is roughly proportional to the process size; therefore, using an FPGA resulting from a smaller process could produce an \ADRS with a very high sampling rate of $>100$ GSa/s.

\begin{ack}

The authors would like to thank to Masayoshi Todorokihara (Seiko Epson), Yuhki Hashimoto (Spedia), and Yasunori Fujii (National Astronomical Observatory of Japan) for their kindly technical advice.
We are grateful to all the staffs, especially for Shigeru Fuji and Ken'ichi Tatematsu, in Nobeyama Radio Observatory for their kindly help to operate the Osaka Prefecture University 1.85-m mm-submm telescope.
The telescope is developed and operated by a lot of contributions of people engaged in this project, so authors would like to thank all the people, especially for Shinpei Nishimoto, Sho Masui, Yasumasa Yamasaki, Hiroshi Kondo, Ayu Konishi, Sana Kawashita, Sho Yoneyama.
This work was supported by JSPS KAKENHI grant numbers,
JP18H05440% Onishi - 新学術
.
This research made use of astropy, a community-developed core Python package for Astronomy \citep{2013A&A...558A..33A}, in addition to NumPy and SciPy \citep{2011CSE....13b..22V}, Matplotlib \citep{2007CSE.....9...90H} and  IPython\citep{2007.ipython}.

\end{ack}

\newpage

% \begin{figure}[t]
%  \begin{center}
%   \includegraphics[width=\textwidth]{fig/fpga.png}
%  \end{center}
% \caption{
% Implementation of the analog-to-digital converter (ADC) and the fast Fourier transform (FFT) cores on the field-programmable gate array (FPGA).
% }
% \label{fig:fpga}
% \end{figure}

\clearpage

% References
\bibliography{main} % bibliography data in report.bib
\bibliographystyle{pasjbib2020} % makes bibtex use spiebib.bst

\end{document}